\documentclass[a4paper]{article}
\usepackage{Odyssey2020}
\usepackage{epsfig,amssymb,amsmath,graphicx,amsfonts,xcolor,url,cite,booktabs,comment,multirow,siunitx}

\ninept

\pdfoutput=1
\newcommand{\spcell}[2][c]{%
  \begin{tabular}[#1]{@{}c@{}}#2\end{tabular}}
  
\newcommand{\frG}{\ensuremath{\mathcal{G}}}
\newcommand{\frR}{\ensuremath{\mathcal{R}}}

\DeclareMathOperator*{\argmax}{arg\,max}

\title{Linguistically Aided Speaker Diarization Using Speaker Role Information}

\name{Nikolaos Flemotomos, Panayiotis Georgiou, Shrikanth Narayanan}

\address{Department of Electrical and Computer Engineering\\ University of Southern California, Los Angeles, CA, USA \\
{\small \tt flemotom@usc.edu, georgiou@ee.usc.edu, shri@ee.usc.edu} }
\begin{document}
\maketitle

\begin{abstract}
Speaker diarization relies on the assumption that speech segments corresponding to a particular speaker 
are concentrated in a specific region of the speaker space; a region which represents that speaker's identity. These identities are not known a priori, so a clustering algorithm is typically employed, which is traditionally based solely on audio. Under noisy conditions, however, such an approach poses the risk of generating unreliable speaker clusters. In this work we aim to utilize linguistic information as a supplemental modality to identify the various speakers in a more robust way. We are focused on conversational scenarios where the speakers assume distinct roles and are expected to follow different linguistic patterns. This distinct linguistic variability can be exploited to help us construct the speaker identities. That way, we are able to boost the diarization performance by converting the clustering task to a classification one. The proposed method is applied in real-world dyadic psychotherapy interactions between a provider and a patient and demonstrated to show improved results.
\end{abstract}
%
\section{Introduction}
Given a speech signal with multiple speakers, diarization answers the question ``who spoke when''~\cite{anguera2012speaker}. To address the problem, the main underlying idea is that speech segments corresponding to some speaker share common characteristics which are ideally unique to the particular person. So, the problem is usually reduced to finding a suitable representation of the signal and a reliable distance metric. Under this viewpoint, when the distance between two speech segments is beyond a certain threshold, they are considered to belong to different speakers.

First, the input signal is segmented either uniformly (e.g.~\cite{sell2018diarization}) or according to a speaker change detection algorithm (e.g.~\cite{zajic2018zcu}). In either case, it is assumed that a single speaker is present in each one of the resulting segments. Since diarization is typically viewed as an unsupervised task, it heavily depends on the successful application of a clustering algorithm in order to group same-speaker segments together. Such a method, however, poses the risk of creating noisy, non-representative speaker clusters. In particular, if the speakers to be clustered reside closely in the speaker space, some speakers may be merged.
Additionally, if there is enough noise and/or silence within a recording (possibly not sufficiently captured by a voice activity detection algorithm), it may be the case that one of the constructed clusters only contains the non-speech or distorted-speech segments. This behavior can lead to poor performance even if the exact number of speakers is known in advance.

Even though speaker diarization has traditionally been an audio-only task which relies on the acoustic variability between different speakers, the linguistic content captured in the speech signal can offer valuable supplementary cues. Apart from practical observations such as the fact that it is highly improbable for a speaker change point to be located within a word~\cite{silovsky2012incorporation}
it is widely accepted that each individual has their very own way of using language~\cite{johnstone1996linguistic}. 
Thus, language patterns followed by individual speakers have been explored in the literature for the tasks of speaker segmentation and clustering, both when used unimodally~\cite{meng2017hierarchical}, and in combination with the speech audio~\cite{india2017lstm,zajic2018recurrent,Park2019}. 

Using language for diarization can be especially promising in structured scenarios where the speakers assume dissimilar roles with distinguishable linguistic patterns. For example, a teacher is likely to speak in a more didactic style while a student be more inquisitive, a doctor is likely to inquire on symptoms and prescribe while a patient describe symptoms, etc. In our own past work~\mbox{\cite{flemotomos2018combined,flemotomos2019role}} we exploited the differences of language between a therapist and a patient to facilitate the automatic processing of medical conversations. The strong interconnection between speech and language in such scenarios has even led to end-to-end role-annotated Automatic Speech Recognition (ASR) architectures~\cite{flemotomos2018role, el2019joint} where diarization becomes a byproduct of a rich transcription system.

Despite the beneficial effects of using language as an additional stream of information, there is an important practical consideration: how to get access to the transcripts. In a real-world scenario, a high-performing ASR system needs to be applied before any textual data is available. However, speaker diarization is widely viewed as a pre-processing step of multi-talker ASR systems and is often a module that precedes ASR in conversational speech processing pipelines~\mbox{\cite{huang2007ibm,xiao2016technology}}. This is because single-speaker speech segments allow for speaker normalization techniques, including 
speaker adaptive training through Constrained Maximum Likelihood Linear Regression (\mbox{CMLLR})~\cite{gales1998maximum} and i-vector based neural network adaptation~\cite{saon2013speaker}. Nevertheless, taking into consideration the error propagating from a non-ideal diarization system to the ASR output, it is nowadays questionable whether diarization can in practice improve recognition accuracy, which is why several modern pipelines start by applying ASR first, achieving state-of-the-art results~\cite{Park2019,yoshioka2019meeting}. In any case, if there are not major computational and/or time constraints, running a second pass of ASR after diarization could be a reasonable approach.

Following the aforementioned line of work, we propose an alternative way of using the linguistic information for the task of speaker diarization in recordings where participants play specific roles which are known in advance. In particular, we process the text stream independently in order to segment it in speaker-homogeneous chunks (where only one speaker is present), each one of which can be assigned to one of the available speaker roles. Aggregating this information for all the segments, and aligning text with audio, we can construct the acoustic identities of the speakers found in the recording. That way, each audio segment can be assigned to a speaker through a simple classifier, overcoming the potential risks of clustering. We apply this approach in psychotherapy recordings featuring dyadic interactions between two speakers with well-defined roles; namely that of a therapist and a patient. 

The rest of the paper is structured as follows: Section~\ref{sec:audio_background} provides a background on audio-based speaker diarization systems and on speaker role recognition in structured conversations. In Section~\ref{sec:method} we present our approach on how to use the linguistic information in order to estimate the speaker identities instead of relying on clustering. In Sections~\ref{sec:data} and~\ref{sec:exp} we give an overview of the datasets used and we analyze our experimental results. Finally, in Section~\ref{sec:conc} we discuss limitations of our work, as well as potential directions for future research.

\section{Background}
\label{sec:audio_background}
\subsection{Audio-Only Speaker Diarization}
Speaker diarization is the process of partitioning a speech signal into speaker-homogeneous segments and then grouping same-speaker segments together, without having prior information about the speaker identities. Therefore, research effort has been focused on finding i) a representation that can capture speaker-specific characteristics and ii) a suitable distance metric that can separate different speakers based on those characteristics. The traditional approach has been to model speech segments under some probability distribution (e.g., Gaussian Mixture Models - GMMs), and measure the distance between them using a metric based on the Bayesian Information Criterion (BIC)~\cite{chen1998speaker}. 

Speaker modelling by GMMs was later replaced by i-vectors \cite{shum2011exploiting}, fixed-dimensional embeddings inspired by the total variability model. In this framework, the cosine distance metric was initially proposed as the divergence criterion to be used, but Probabilistic Linear Discriminant Analysis (PLDA) based scoring \cite{ioffe2007probabilistic,prince2007probabilistic} was proved to yield improved results \cite{sell2014speaker}. Given two embeddings $v$, $r$, PLDA provides a framework to estimate their similarity $s(v,r)$ as the log-likelihood ratio 
\begin{equation}\label{eq:plda}
s(v,r) = \log\frac{p(v,r|\text{same speaker})}{p(v|\text{dif. speakers})p(r|\text{dif. speakers})}
\end{equation}

In recent years, with the advent of Deep Neural Networks (DNNs), the embeddings used are usually bottleneck features extracted from neural architectures. Such architectures are trained under the objective of speaker classification employing a cross-entropy loss function~\cite{snyder2018x}, or speaker discrimination employing contrastive~\cite{garcia2017speaker} and triplet~\cite{bredin2017tristounet} loss functions. Typical examples of embeddings that have shown state-of-the-art performance for speaker diarization are the Time-Delay Neural Net (TDNN) based x-vectors~\cite{sell2018diarization} and the Long-Short Term Memory (LSTM) based d-vectors~\cite{wang2018speaker}. 

Speaker diarization usually comprises two steps: first, the speech signal is segmented into single-speaker chunks, and second, the resulting segments are clustered into same-speaker groups~\cite{anguera2012speaker}. Even though speaker change detection is by itself an active research field~\mbox{\cite{hruz2017convolutional,jati2017speaker2vec}}, it has been shown that it doesn't necessarily lead to improved results within the framework of diarization when compared to a uniform, sliding-window based segmentation~\mbox{\cite{zajic2016investigation,zajic2018zcu}}, so the latter method is widely used. As far as the clustering is concerned, common approaches include Hierarchical Agglomerative Clustering (HAC)~\cite{sell2018diarization} and spectral clustering~\cite{wang2018speaker,park2019second}, while methods based on affinity propagation~\cite{yin2018neural} and generative adversarial networks~\cite{pal2019speaker} have also been proposed. In order to overcome some of the problems connected with clustering, supervised systems that directly output a sequence of speaker labels have been recently introduced~\mbox{\cite{zhang2019fully,fujita2019end}}.

\subsection{Speaker Role Recognition}
In many real-world speech applications speakers play specific roles, in the sense that they perform well-defined tasks within the conversation domain. For example, we can think of the different speakers who may appear during a broadcast news program, a clinical visit, or a lecture. Typically, the domain is given, so the set of available roles is known in advance and Speaker Role Recognition (SRR) is viewed as a supervised classification task~\mbox{\cite{barzilay2000rules,bigot2013speaker}}. Even though various modalities have been explored in the literature~\cite{salamin2011automatic,rouvier2015multimodal,silber2020positioning}, the linguistic variability between the speaker roles usually provides the most important cues for the task in hand~\cite{flemotomos2019role}.

The lexical information was initially captured by n-gram features which were passed to boosting algorithms or maximum entropy classifiers~\cite{barzilay2000rules}. Additional language-based features that have been proposed include the types of questions posed by different speakers~\cite{bazillon2011speaker}, as well as the errors identified in ASR transcripts~\cite{damnati2011robust}. Deep learning approaches have been explored in~\cite{rouvier2015multimodal} where SRR relies on word embeddings and Convolutional Neural Networks (CNNs). In~\cite{xiao2016technology} SRR is viewed as a problem of finding the role-specific n-gram Language Model (LM) which minimizes the perplexity of a text segment, while in~\cite{flemotomos2019role} the role assignment is based on the LM scores of role-adapted ASR systems.

\section{Linguistically-Aided Speaker Diarization}
\label{sec:method} 
\begin{figure}[ht]
  \centering
  \includegraphics[width=\columnwidth]{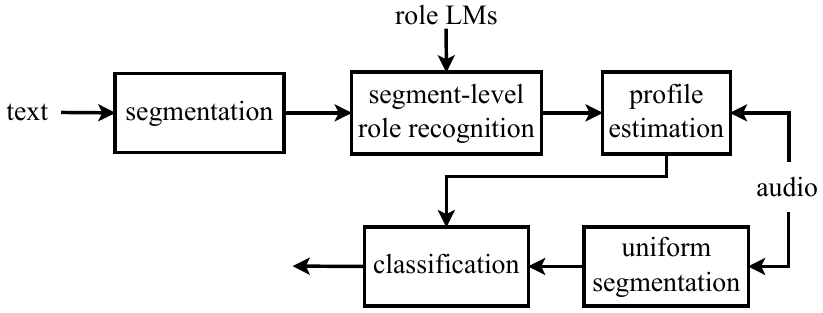}
  \caption{\ninept Linguistically-aided speaker diarization using role information.}
  \label{fig:proposed}
\end{figure}

Our proposed approach for speaker diarization in conversational interactions where speakers assume specific roles is illustrated in Figure~\ref{fig:proposed}.
We describe the various modules in detail in Sections \ref{subsec:segmenter}-\ref{sec:dia_w_classification}.

\subsection{Text-based Segmentation}
\label{subsec:segmenter}
Given the textual information of the conversation, our goal is to obtain speaker-homogeneous text segments; that is segments where all the words have been uttered by a single speaker. Those will later help us construct the desired acoustic speaker identities. 
Even though text-based speaker change detectors have been proposed~\cite{meng2017hierarchical}, for our final goal we can safely over segment the available document, provided this leads to a smaller number of segments containing more than one speakers~\cite{zajic2018recurrent}. So, we assume that each sentence is with high probability speaker-homogeneous and we instead segment at the sentence level.

\begin{figure}[ht]
  \centering
  \includegraphics[width=.55\columnwidth]{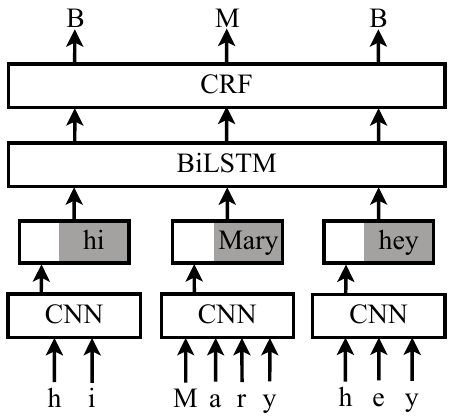}
  \caption{\ninept Neural network for sentence-level text segmentation: A character representation is constructed for each word through a CNN and is concatenated with a word embedding (here shown in grey). This is the input to a BiLSTM-CRF architecture which predicts a sequence of labels. Here `B' denotes a word at the beginning of a sentence and `M' in the middle (any word which is not the first one of a sentence).}
  \label{fig:tag_nn}
\end{figure}

To that end, the problem can be viewed as a sequence labeling one, where each word is tagged as either being at the beginning of a sentence, or anywhere else. 
In particular, we address the problem building a Bidirectional LSTM (BiLSTM) network with a Conditional Random Field (CRF) inference layer~\cite{ma2016end}, as shown in Figure~\ref{fig:tag_nn}. The input to the recurrent layers is a sequence of words. Each word is given as a concatenation of a character-level representation predicted by a CNN and a word embedding. For our experiments, we initialize the word embeddings with the extended dependency skip-gram embeddings~\cite{komninos2016dependency}, pre-trained on $2$B words of Wikipedia. Those extend the semantic vector space representation of the word2vec model~\cite{mikolov2013distributed} considering not only spacial co-occurrences of words within text, but also co-occurrences in a dependency parse graph. That way, they can capture both functional and topic related semantic properties of words.

\subsection{Role Recognition}
\label{subsec:srr}
The next step in our system is the application of a text-based role recognition module. In more detail, assuming we have $N$ speakers in the session ($N=2$ for our experiments) and there is one-to-one correspondence between speakers and roles (e.g., there is one therapist and one patient), we want to assign one of the role labels $\{R_i\}_{i=1}^N$ to each segment. To do so, we build $N$ LMs $\{\mathcal{R}^+_i\}_{i=1}^N$, one for each role, and we estimate the perplexity of a segment given the LM $\mathcal{R}^+_i$, for $i=1,2,\cdots,N$. The role assigned to the segment is the one yielding the minimum perplexity~\cite{flemotomos2018combined}. We note that all the perplexities are normalized for segment length. 

The required role-specific LMs are n-gram models built as described in~\cite{flemotomos2019role}. For this process, we assume that in-domain text data is available for training. We first construct a background, out-of-domain LM $\frG$ and $N$ role-specific LMs $\{\mathcal{R}_i\}_{i=1}^N$. $\frG$ is used to ensure a large enough vocabulary that minimizes the unseen words during the test phase. Those individual LMs are interpolated to get the mixed models $\{\mathcal{R}^+_i\}_{i=1}^N$. To obtain an interpolated model, the weighted average of the probabilities from the input models is assigned to each n-gram, and the produced model is then re-normalized~\cite{stolcke2002srilm}. Denoting LM interpolation by the symbol $\oplus$, the final models are
\begin{eqnarray}
&\frG^+ = w_g\frG\oplus (1-w_g)\tilde{\frR} \label{eq:lm_g}\\
&\frR_{i}^+ = w_{g_i}\frG\oplus w_{r_i}\frR_i \oplus (1-w_{g_i}-w_{r_i})\tilde{\frR}_i \label{eq:lm_r}
\end{eqnarray}
where
$$\tilde{\frR} = \frac{1}{N}\bigoplus_{i=1}^N\frR_i,\;\;\text{}\;\;\tilde{\frR}_i = \frac{1}{N-1}\bigoplus_{\substack{j=1\\j\neq i}}^N\frR_j$$
and all the weights $w_g, w_{g_i}, w_{r_i}$ are chosen to minimize the perplexity of appropriate role-specific development corpora.

\subsection{Profile Estimation}
\label{subsec:profiles}
After the application of the text-based segmenter and role recognizer, we have several text segments corresponding to each role $R_i$. If we have the alignment information at the word level\footnote{If we have access to the transcripts and the audio, we can force-align. If we generate the text through ASR, we get the desired alignments from the decoding lattices.}, we can directly get the time-boundaries of those segments. We extract one embedding (x-vector) for each and we estimate a role identity $r_i$ as the mean of all the embeddings corresponding to the specific role. Under the assumption of one-to-one correspondence between speakers and roles that we already introduced in Section~\ref{subsec:srr}, those role identities are at the same time the acoustic identities (also known as profiles) of the \emph{speakers} appearing in the initial recording. 

We note that the role recognition at the segment level does not always provide robust results~\cite{flemotomos2019role}, something which could lead to unreliable generated profiles. However, we expect that there will be a fraction of the segments for the results of which we are confident enough and we can take only those into consideration for the final averaging. The proxy used as our confidence for the segment-level role assignment is the difference between the best and the second best perplexity of a segment given the various LMs. Formally, if the segment $x$ is assigned the role $R_i$, and if $pp(x|\mathcal{R}^+_i)$ is the perplexity of $x$ given the LM $\mathcal{R}^+_i$, then the confidence metric used for this assignment is 
$c_x = \min_{j\neq i}| pp(x|\mathcal{R}^+_j) - pp(x|\mathcal{R}^+_i) |$. Then, the corresponding profile is
\begin{equation}\label{eq:theta}
r_i = \frac{\sum_{x\in R_i}\tilde{c}_xu_x}{\sum_{x\in R_i}\tilde{c}_x} \triangleq \frac{\sum_{x\in R_i}\mathbb{I}\{c_x>\theta\}u_x}{\sum_{x\in R_i}\mathbb{I}\{c_x>\theta\}}
\end{equation}
where $u_x$ is the x-vector representing the segment $x$, $\mathbb{I}\{\cdot\}$ is the indicator function, and $\theta$ is a tunable parameter. 

\subsection{Audio Segmentation and Classification}
\label{sec:dia_w_classification}
After having computed all the needed profiles $\{r_i\}_{i=1}^N$, in order to perform speaker diarization, we first segment the audio stream of the speech signal uniformly with a short sliding window, a typical approach in audio-only diarization systems. In other words, the language information is used by our framework only to construct the speaker profiles, with the final diarization result relying on audio-based segmentation, as illustrated in Figure~\ref{fig:proposed}. For each one of the resulting segments an \mbox{x-vector} is extracted. However, instead of clustering the \mbox{x-vectors}, we now classify them within the correct speaker/role. In order to have a fair comparison between common diarization baselines and our proposed system, our classifier is based on PLDA, but we note that any other classifier could be employed instead. In this framework, a segment $x$ with embedding $u_x$ is assigned the label
$\hat{R}_x = \argmax_{1\leq i\leq N}\{s(u_x,r_i)\}$, where $s(\cdot,\cdot)$ is the PLDA similarity score estimated in equation~\eqref{eq:plda}.

\section{Datasets}
\label{sec:data}
\subsection{Evaluation Data}
We evaluate our proposed method on datasets from the clinical psychology domain. In particular, we apply the system to motivational interviewing sessions -- a specific type of psychotherapy -- between a therapist and a patient, collected from five independent clinical trials (ARC, ESPSB, ESP21, iCHAMP, HMCBI) \cite{atkins2014scaling}. We collectively refer to those sessions as the PSYCH corpus. This is split into training, development, and test sets, as shown in Table~\ref{table:data_MI}, in such a way that there is no speaker overlap between them. Even though the necessary metadata is available for the rest of the corpus, the patient IDs are not available for the HMCBI sessions. Thus, the partitioning is done under the assumption that it is highly improbable for the same patient to visit different therapists in the same study. All the results reported are on PSYCH-test.

\begin{table}[ht]
  \centering
	\begin{tabular}{c|c|c|c}
    	\toprule
    	& PSYCH-train & PSYCH-dev & PSYCH-test \\
        \midrule
        \#sessions & $74$ & $44$ & $25$\\
        \midrule
        Therapist& \SI{26.43}{h} & \SI{15.23}{h} & \SI{7.34}{h}\\
        Patient& \SI{23.29}{h} & \SI{12.17}{h} & \SI{7.54}{h}\\
       	\bottomrule
	\end{tabular}    
	\caption{\ninept Size of the PSYCH dataset. The total durations of the speech segments corresponding to a therapist and to a patient are calculated based on the manual turn boundaries.}
	\label{table:data_MI}
\end{table}

\subsection{Segmenter and Role LM Training Data}
The segmenter presented in Section \ref{subsec:segmenter} is trained on a subset of the Fisher English corpus \cite{cieri2004fisher} comprising a total of $10$,$195$ telephone conversations for which the original transcriptions (including punctuation symbols which are essential for the training of our network) are available. This set is enhanced by $1$,$199$ in-domain therapy sessions provided by the Counseling and Psychotherapy Transcripts Series\footnote{\url{https://alexanderstreet.com/products/counseling-and-psychotherapy-transcripts-series}} (CPTS). The combined dataset is randomly split ($80$-$20$ split at the session level) into training and validation sets. CPTS and the entire Fisher English corpus, together with the training part of the PSYCH corpus, are also used to train the required role-specific LMs. The size of the corresponding vocabularies is given in Table \ref{table:voc_size}.

\begin{table}[ht]
  \centering
	\begin{tabular}{c|c|c|c}
    	\toprule
    	& PSYCH-train & Fisher & CPTS \\
        \midrule
        $|$voc$|$& $8.17$K & $58.6$K & $35.6$K \\
        \#tokens& $530$K & $21.0$M & $6.52$M \\
        \bottomrule
	\end{tabular}    
	\caption{\ninept Size of the vocabulary and total number of tokens in the corpora used for LM training.}
	\label{table:voc_size}
\end{table}

\section{Experiments and Results} \label{sec:exp}
\subsection{Baseline Systems}
\label{subsec:baseline}
\subsubsection{Audio-based Diarization with Speaker Clustering}
\label{subsubsec:baseline_audio}
\begin{figure}[ht]
  \centering
  \includegraphics[width=.95\columnwidth]{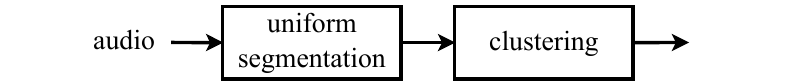}
  \caption{\ninept Baseline audio-based speaker diarization.}
  \label{fig:baseline}
\end{figure}

As an audio-only baseline, we use a state-of-the-art speaker diarization system following the x-vector/PLDA paradigm~\cite{sell2018diarization}. As shown in Figure~\ref{fig:baseline}, the speech signal is first segmented uniformly and an x-vector is extracted for each segment. The pairwise similarities $s(\cdot,\cdot)$ between all those embeddings are then calculated based on PLDA scoring (equation \eqref{eq:plda}).

The segments are clustered into same-speaker groups following a HAC approach with average linking. Since our experiments are conducted on dyadic interactions, we force the HAC algorithm to run until two clusters are constructed. 

\subsubsection{Language-based Diarization}
\label{subsubsec:baseline_lang}
\begin{figure}[ht]
  \centering
  \includegraphics[width=\columnwidth]{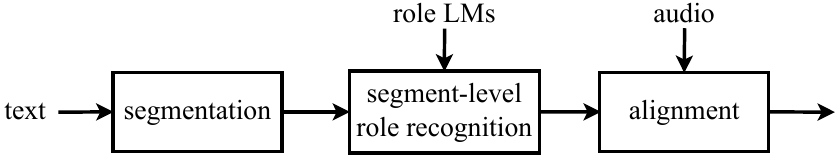}
  \caption{\ninept Baseline language-based speaker diarization.}
  \label{fig:baseline_lang}
\end{figure}

As a language-only baseline, we use the system of Figure~\ref{fig:baseline_lang}, which essentially consists of the first steps of the framework in Figure~\ref{fig:proposed}. After estimating the segment-level role labels as described in Sections~\ref{subsec:segmenter} and~\ref{subsec:srr}, we can simply use those as our diarization output labels to compare the performance of a system that only depends on the linguistic information. In that case, we only utilize audio to get the timestamps of the text segments. If an ASR system is used, this information is already available through the decoding lattices.

\subsection{Experimental Setup}
\label{subsec:exp}
As a pre-processing step, the text which is available from the manual transcriptions of the PSYCH corpus is normalized to remove punctuation symbols and capital letters, and force-aligned with the corresponding audio sessions. Based on the word alignments, we segment the audio according to whether there is a silence gap between two words larger than a threshold equal to $\SI{1}{sec}$. We should highlight that this initial segmentation is applied before running either one of the baseline systems or our proposed architecture. Thus, the initial segments to be diarized are always the same and those are also the segments that we pass to the ASR system. The diarization ground truth is also constructed through the word alignments, by allowing a maximum of $\SI{0.2}{sec}$-long in-turn silence.

The resulting text segments are further subsegmented at the sentence level based on the output of the tagger in Figure~\ref{fig:tag_nn}. During training we define as ``sentence'' any text segment between two punctuation symbols denoting pause apart from commas. We exclude commas first because they normally do not indicate speaker change points but also because they are too frequent in our training set and they would lead to very short segments, not containing sufficient information for the task of role recognition. The tagger is built using the NCRF++ toolkit~\cite{yang2018ncrf}. Following the general recommendations in~\cite{reimers2017reporting} and after our own hyperparameter tuning, the network comprises $4$ CNN layers and $2$ stacked BiLSTM layers with dropout ($p=0.5$) and $l_2$ regularization ($\lambda=10^{-8}$). The length of each word representation is 330 (character embedding dimension = $30$, word embedding dimension = $300$). The network is trained using the Adam optimizer with a fixed learning rate equal to $10^{-3}$ and a batch size equal to $256$ word sequences. The tagger achieves an $F_1$ score of $0.805$ on the validation set after $14$ training epochs. 

All the LMs required for role recognition are 3-gram models with Kneser-Ney smoothing built with the SRILM toolkit~\cite{stolcke2002srilm}. LMs are trained using a combination of in-domain data from the PSYCH-train and CPTS corpora and out-of-domain data from the Fisher English corpus, employing the LM interpolation procedure described in Section~\ref{subsec:srr}, with mixing weights optimized on PSYCH-dev. SRILM is also used to estimate the needed perplexities. 

The audio-based diarization framework is built using the Kaldi toolkit~\cite{povey2011kaldi}. Both for the audio-based baseline and for our system, we use the VoxCeleb pre-trained x-vector extractor\footnote{\url{https://kaldi-asr.org/models/m7}} and the PLDA model which comes with it, after we adapt it on the development set of the PSYCH corpus. The x-vectors are extracted after uniformly segmenting the audio into \SI{1.5}{sec}-long windows with $50\%$ overlap. Those are normalized and decorrelated through an LDA projection and dimensionality reduction (final embedding length = 200), mean, and length normalization. The evaluation is always based on the Diarization Error Rate (DER), as estimated by the NIST \texttt{md-eval.pl} tool, with a \SI{0.25}{sec}-long collar, ignoring overlapping speech.

To get the ASR outputs, we use Kaldi’s pre-trained \mbox{ASpIRE} acoustic model\footnote{\url{http://kaldi-asr.org/models/m1}}, coupled with the n-gram LM given in equation~\eqref{eq:lm_g}. This ASR system gives a Word Error Rate (WER) equal to $38.02\%$ for the PSYCH-dev and $39.78\%$ for the PSYCH-test set. It is noted that WERs in this range are typical in spontaneous medical conversations~\cite{kodish2018systematic}.

\subsection{Results with Reference Transcripts}
\label{subsec:exp_ref}
Before applying ASR, we employ our system using the manually derived transcripts. That way, we can inspect the usability and effectiveness of our idea, eliminating potential propagation errors because of ASR. Table~\ref{table:res_ref} gives the results of our linguistically-aided diarization system in comparison with the audio-only and language-only baseline approaches. First, we notice that, between the two baselines, the one using the acoustic modality yields better results. This came at no surprise since we expected that audio carries the most important speaker-specific characteristics. Hence, we propose using language only as a supplementary stream of information. 

\begin{table}[ht]
  \centering
	\begin{tabular}{c||c|c|c}
    	\toprule
    	\spcell{text\\segmentation}& \spcell{audio\\only} & \spcell{language\\only} & \spcell{linguistically\\aided}\\
    	\midrule
    	oracle & \multirow{2}{.8cm}{$11.05$} & $12.99$ & $7.28$\\
    	tagger &  & $20.09$ & $7.71$ \\
        \bottomrule
	\end{tabular}    
	\caption{\ninept DER $(\%)$ following our linguistically-aided approach and the two baselines. The text segmentation (when needed) is either performed by our sequence \emph{tagger} or based on the manually annotated speaker changes (\emph{oracle}).}
	\label{table:res_ref}
\end{table}

When we apply our linguistically-aided system using our sequence tagger to segment at the sentence level we get a $30.23\%$ DER relative improvement compared to the audio-only approach. In the first row of Table~\ref{table:res_ref} we additionally report results when using the oracle speaker segmentation provided by the manual annotations instead of applying the sequence tagger. That way, we can eliminate any negative effects caused by a suboptimal speaker change detector. As expected, the results are indeed better, but it is worth noting the difference in the performance gap between the language-only and the linguistically-aided approaches. Since the sequence tagger operates at the sentence level, its output is over-segmented with respect to speaker changes. As a result, utterances are broken into very short segments, with several segments containing insufficient information to infer speaker role in a robust way.
However, when we aggregate all those speaker turns to only estimate an average speaker profile, such inaccuracies cancel out.

Further improvements are observed if we only keep the segments we are most confident about for the profile estimation, applying the confidence metric introduced in Section~\ref{subsec:profiles}. Instead of directly optimizing for the parameter $\theta$ appearing in equation~\eqref{eq:theta}, we find the parameter $a$ that minimizes the overall DER on the development set when only the $a\%$ segments we are most confident about are taken into consideration per session. The results on the test set are reported in Table~\ref{table:res_dagger}. An additional $5.32\%$ relative error reduction is achieved when our tagger is used and similar improvements are noticed in the case of oracle text segmentation.

\begin{table}[ht]
  \centering
	\begin{tabular}{c||c|c}
    	\toprule
    	\spcell{text segmentation}& DER $(\%)$ & DER$^\dagger$ $(\%)$\\
    	\midrule
    	oracle & $7.28$ & $6.99$\\
    	tagger & $7.71$ & $7.30$ \\
        \bottomrule
	\end{tabular}    
	\caption{\ninept Diarization results following our linguistically-aided approach. The text segmentation is either performed by our sequence \emph{tagger} or based on the manually annotated speaker changes (\emph{oracle}). $\dagger$ denotes results when only $a\%$ of the segments we are most confident about are taken into account in each session for the profile estimation, where $a$ is a parameter optimized on the development set ($a=70$ for the tagger segmentation and $a=55$ for the oracle segmentation).}
	\label{table:res_dagger}
\end{table}

\begin{figure}[ht]
  \centering
  \includegraphics[width=.9\columnwidth]{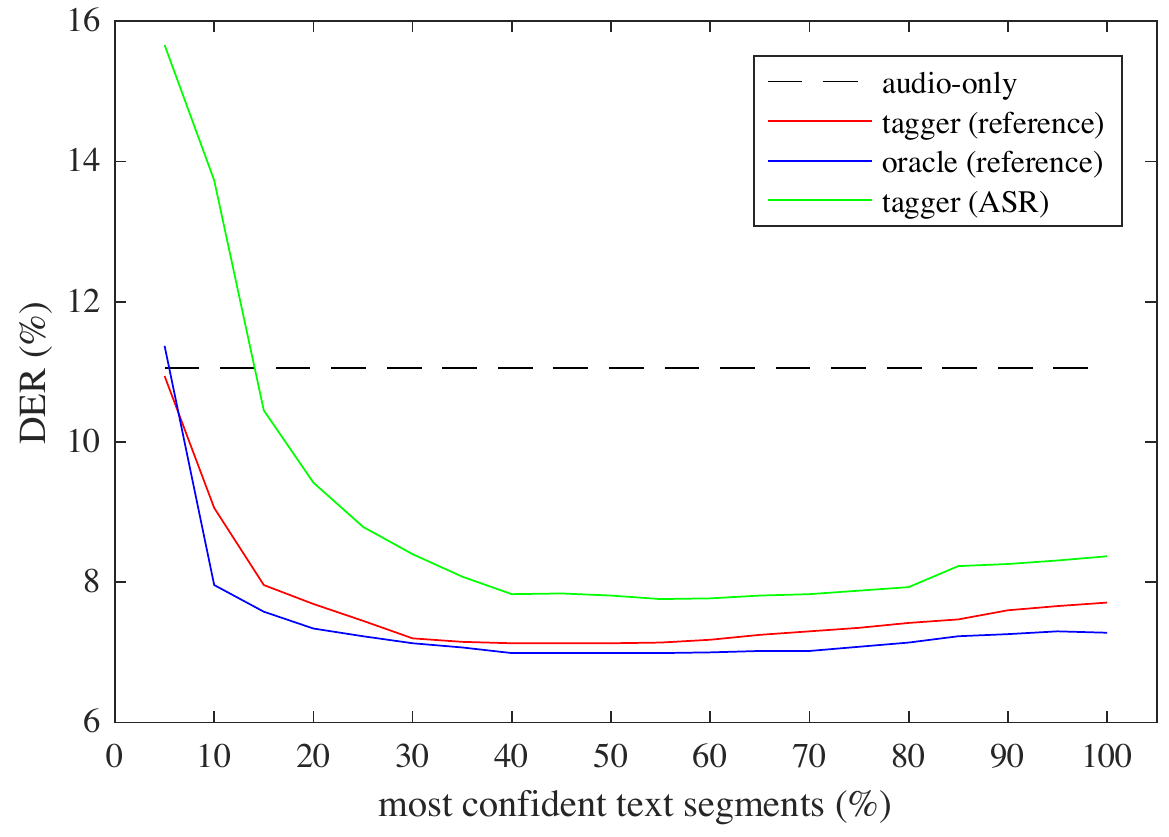}
  \caption{\ninept DER $(\%)$ as a function of the percentage of the text segments we take into account per session for the profile estimation, based on our confidence metric. The text segmentation is either performed by our sequence \emph{tagger} or based on the manually annotated speaker changes (\emph{oracle}). Results are presented both with \emph{reference} and with \emph{ASR} transcripts.}
  \label{fig:der_frac}
\end{figure}

In Figure~\ref{fig:der_frac} we plot DER as a function of the percentage of the segments we use to estimate the speaker profiles within a session. Even though the oracle text segmentation consistently yields marginally better results, it seems that if we carefully choose which segments to use to get an estimate of the speakers' identities, our tagger-based segmentation approaches the oracle performance. In fact, the best result we got on the test set (optimizing for $a$ on the same set) using our segmenter was $7.13\%$ DER, while the corresponding number using the oracle segmentation was $6.99\%$. We should highlight here that the analysis presented in this work is based on using $a\%$ of the segments within a session, after choosing some $a$ which remains constant across sessions. It is probable that this is a session-specific parameter which ideally should be chosen based on an alternative, session-level strategy.

\subsection{Results with ASR Transcripts}
\label{subsec:exp_asr}
For the experiments in this Section we apply the same pre-processing steps, but we replace the reference transcripts with the textual outputs of the ASR system and the corresponding time alignments. The results are given in Table~\ref{table:res_asr}. Here, we report results only when using the sequence tagger (and not with oracle segmentation). The reason is simply because we now assume we have no access to the reference transcripts, so we cannot know the oracle speaker change points. 

\begin{table}[ht]
  \centering
	\begin{tabular}{c|c|c|c}
    	\toprule
    	\spcell{method}& \spcell{audio\\only} & \spcell{language\\only} & \spcell{linguistically\\aided}\\
    	\midrule
    	DER $(\%)$ & $11.05$ & $27.07$ & $8.37$ \\
        \bottomrule
	\end{tabular}    
	\caption{\ninept Diarization results following our linguistically-aided approach and the two baselines. The text segmentation (when needed) is performed by our sequence tagger.}
	\label{table:res_asr}
\end{table}

As we can see, the diarization performance is substantially improved compared to the audio-only system (relative DER reduction equal to $24.25\%$) even if the WER of the ASR module is relatively high, as reported in Section~\ref{subsec:exp}. It seems that when using the transcripts only for the task of profile estimation, the overall performance is not severely degraded by a somehow inaccurate ASR system. As expected, this is not the case for our language-only baseline. Since in that case the final output only depends on the linguistic information, the performance gap between using manually- and ASR-derived transcripts (2nd column in Tables~\ref{table:res_ref} and~\ref{table:res_asr}) is large. We should note that this performance gap is not only due to higher speaker confusion in the case of ASR transcripts, but also because of increased missed speech. In particular, the missed speech when using ASR is $2.7\%$ because of word deletions (as opposed to $0.6\%$ when the reference transcrpits and the tagger are used).

As was the case with the experiments in Section~\ref{subsec:exp_ref}, further improvements are observed when only using a subset of the total number of segments per session to estimate the speaker profiles. In particular, if $a=45\%$ of the segments for which we are most confident about (after optimizing for $a$ on the development set) are used, DER is reduced to $7.84\%$. The beneficial effects of using our confidence metric to estimate a speaker representation only by a subset of their assigned speech segments is also demonstrated in Figure~\ref{fig:der_frac}.

\section{Conclusions and Future Work}
\label{sec:conc}
We proposed a system for speaker diarization suitable to use in conversations where the participants assume specific roles, associated with distinct linguistic patterns. While this task typically relies on clustering methods which can lead to noisy speaker partitions, we demonstrated how we can exploit the lexical information captured within the speech signal in order to estimate the speaker profiles and follow a classification approach instead. A text-based speaker change detector is an essential component of our system. For this subtask, assuming each sentence is speaker-homogeneous, we proposed using a sequence tagger which can segment at the sentence level, by detecting the beginning of a new sentence and we showed that this segmentation strategy approaches the oracle performance. The resulting segments are assigned a speaker role label which is later used to construct the desired speaker identities and we introduced a confidence metric to be associated with this assignment. Our results showed that such a metric can be used in order to take into consideration only the segments we are most confident about, leading to further performance improvements. When applied to dyadic interactions between a therapist and a patient, our proposed method achieved an overall relative DER reduction equal to $29.05\%$, compared to the baseline audio-only approach with speaker clustering. When reference transcripts were used instead of ASR outputs, the corresponding overall reduction was equal to $33.94\%$.

Since role recognition is a supervised task, one drawback of our system when compared to traditional diarization approaches is that it requires in-domain text data in order to build the role-specific LMs. It should be additionally highlighted that the diarization results can be further improved if, for example, a re-segmentation module is employed as a final step, or a more precise audio segmentation strategy is followed instead of relying on uniform segmentation. For instance, an audio-based speaker change detector could be applied both for the audio-only baseline and the linguistically-aided system and in the latter case this could be used in combination with the language-based segmenter. However, our goal in this work was mainly to demonstrate the effectiveness of constructing the speaker profiles within a session to be diarized in order to convert the clustering task into a classification one and not to achieve the best possible diarization performance. Additionally, since the initial segmentation was the same both for our system and our audio-only baseline, we expect that any improvements with respect to that part (i.e. more sophisticated segmentation and/or application of re-segmentation techniques) would lead to similar relative improvements to both systems.

Here we essentially modelled each speaker by a single embedding, since for the final profile estimation we averaged over all the speech segments assigned to the corresponding speaker. A potential extension of the current work would be an exploration of alternative speaker identity construction strategies, e.g., representing a speaker by a distribution of embeddings. This is particularly promising in scenarios where the recordings are long enough so that they may incorporate various acoustic conditions or different speaking styles corresponding to the same speaker. Finally, another direction of future work could be towards an in-depth investigation of the confidence metric used for the text-based role assignments and, accordingly, the number of those segments that we should take into consideration when estimating the speaker profiles. Even though in this work we employed a metric purely based on perplexity differences, it is possible that other attributes (e.g., segment length) provide additional information.

\section{Acknowledgements}
This work was supported by the National Institutes of Health (NIH). 

\vfill\pagebreak
\ninept
\bibliographystyle{IEEEbib}
\bibliography{refs}

\end{document}